\documentclass[final,3p,times,twocolumn]{elsarticle}

\usepackage{amssymb}
\usepackage{amsmath}
\usepackage{times}
\usepackage{latexsym}
\usepackage{booktabs}
\usepackage{cleveref}
\usepackage{CJKutf8}
\usepackage{enumitem}
\usepackage{makecell}
\usepackage{multirow}
\usepackage{xcolor, color}
\usepackage{helvet}
\usepackage{courier}
\usepackage{caption}
\newcommand{\newparagraph}[1]{\vspace{2pt} \noindent \textbf{#1}}
\usepackage{graphicx}
\usepackage{caption}
\usepackage{subfigure}

\journal{arxiv}

\begin{document}

\begin{frontmatter}

%% Title, authors and addresses

%% use the tnoteref command within \title for footnotes;
%% use the tnotetext command for theassociated footnote;
%% use the fnref command within \author or \address for footnotes;
%% use the fntext command for theassociated footnote;
%% use the corref command within \author for corresponding author footnotes;
%% use the cortext command for theassociated footnote;
%% use the ead command for the email address,
%% and the form \ead[url] for the home page:
%% \title{Title\tnoteref{label1}}
%% \tnotetext[label1]{}
%% \author{Name\corref{cor1}\fnref{label2}}
%% \ead{email address}
%% \ead[url]{home page}
%% \fntext[label2]{}
%% \cortext[cor1]{}
%% \affiliation{organization={},
%%             addressline={},
%%             city={},
%%             postcode={},
%%             state={},
%%             country={}}
%% \fntext[label3]{}

\title{Data-Augmented and Retrieval-Augmented Context Enrichment in Chinese Media Bias Detection
}

%% use optional labels to link authors explicitly to addresses:
%% \author[label1,label2]{}
%% \affiliation[label1]{organization={},
%%             addressline={},
%%             city={},
%%             postcode={},
%%             state={},
%%             country={}}
%%
%% \affiliation[label2]{organization={},
%%             addressline={},
%%             city={},
%%             postcode={},
%%             state={},
%%             country={}}

\author[a]{Luyang Lin}
\author[c]{Jing Li \corref{cor1}}
\author[a]{Kam-Fai Wong}
\affiliation[a]{organization={The Chinese University of Hong Kong}}
\affiliation[c]{organization={The Hong Kong Polytechnic University}}
\cortext[cor1]{Corresponding author}

% \affiliation{organization={The Chinese University of Hong Kong},%Department and Organization
%             addressline={}, 
%             city={},
%             postcode={}, 
%             state={},
%             country={}}

\begin{abstract}
%% Text of abstract

\textit{\textbf{Warning}: This paper contains content that may be offensive or controversial.}

With the increasing pursuit of objective reports, automatically understanding media bias has drawn more attention in recent research.
However, most of the previous work examines media bias from Western ideology, such as the left and right in the political spectrum, which is not applicable to Chinese outlets.
Based on the previous lexical bias and informational bias structure, we refine it from the Chinese perspective and go one step further to craft data with 7 fine-grained labels.
To be specific, we first construct a dataset with Chinese news reports about COVID-19 which is annotated by our newly designed system, and then conduct substantial experiments on it to detect media bias.
However, the scale of the annotated data is not enough for the latest deep-learning technology, and the cost of human annotation in media bias, which needs a lot of professional knowledge, is too expensive.
Thus, we explore some context enrichment methods to automatically improve these problems.
In Data-Augmented Context Enrichment (DACE), we enlarge the training data; while in Retrieval-Augmented Context Enrichment (RACE), we improve information retrieval methods to select valuable information and integrate it into our models to better understand bias.
Extensive experiments are conducted on both our dataset and an English dataset BASIL.
Our results show that both methods outperform our baselines, while the RACE methods are more efficient and have more potential.
% Finally, we also discuss the future development of media bias detection tasks in the era of large language models.
% \footnote{We will make our annotated corpus publicly available upon publication.}

\end{abstract}

\onecolumn

%%Graphical abstract
\begin{graphicalabstract}
\end{graphicalabstract}

%%Research highlights
\begin{highlights}
\item Research highlights 1:\\
Refine the media bias annotation structure for the Chinese media bias detection task, then contribute a Chinese health field media bias dataset and its benchmark in a multi-label setting.
\item Research highlights 2:\\
We finetune the pre-trained models by the media bias dataset and comprehensively study the difference in the detection of lexical bias and informational bias.
\item Research highlights 3:\\
Propose Data-Augmented and Retrieval-Augmented Context Enrichment methods for media bias detection, and improve the performance of this task.
\end{highlights}

\begin{keyword}
%% keywords here, in the form: keyword \sep keyword
%% PACS codes here, in the form: \PACS code \sep code
%% MSC codes here, in the form: \MSC code \sep code
%% or \MSC[2008] code \sep code (2000 is the default)
Chinese Media Bias Detection \sep Natural Language Processing \sep Data Augmentation \sep Information Retrieval 
% \sep Knowledge Distillation
\end{keyword}

\end{frontmatter}

%% \linenumbers

%% main text
% \section{}
% \label{}
% \section{Introduction}
\section{Introduction}
\label{sec:introduction}

Media bias is a common journalism phenomenon in news articles, where authors inject their personal viewpoints into the reports. Automatic detection of media bias has gained more attention in recent years due to its critical role in pursuing objective information dissemination \cite{streckfuss1990objectivity, schudson2001objectivity}.
However, much of the existing research has been conducted within the context of Western ideology context \cite{hanitzsch2019journalism}, such as the left and right in the political spectrum \cite{iyyer2014political,gangula2019detecting,chen2020analyzing,baly2020we,aksenov2021fine,groeling2013media}.

While from the Chinese perspective, the problem is more complex. The most direct phenomenon is that language expression habits cannot be summarized according to the leaning of stances as Western countries do.
In this situation, the political spectrum may not be the only reason for the formation of media bias \cite{yin2008beyond}, other factors, such as the capital market, also play a crucial role \cite{iwabuchi2010globalization}.
%When the stances are put aside, we focus more on some bias types in language angle, like exaggeration or understatement, which commonly appear in reports in order to gain more attention from readers.

Therefore, we propose a new label system for Chinese media bias.
Considering the limitations of most existing studies, which coarsely characterize media bias with either lexical \cite{yano2010shedding,hamborg2020media,spinde2021automated,krieger2022domain} or informational features \cite{card2015media,baumer2015testing,morstatter2018identifying,budak2016fair}, we align with the concept proposed by Fan et al \cite{fan2019plain}, which consists both \textbf{lexical bias} and \textbf{informational bias}.
% In this frame, it can not only show how a news excerpt is written but also what content is selected to focus on. 
Further, to define the bias more clearly and objectively, we craft 7 fine-grained labels following the journalism theories \cite{baker1994identify} and discuss with domain experts in media and journalism.

\begin{figure}[htp]
    \centering
    \includegraphics[width=0.45\textwidth]{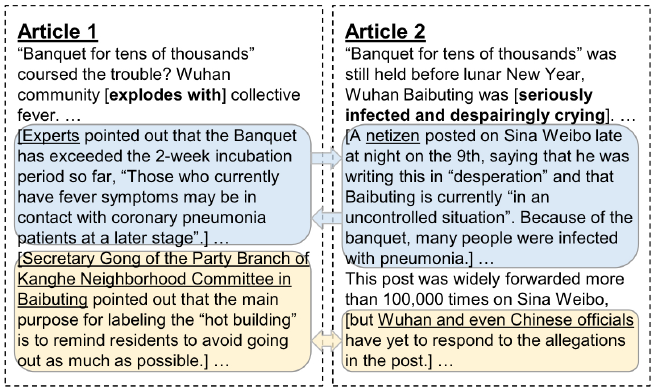}
    % \vspace{-0.8em}
    \caption{
    Two COVID-19 news excerpts from media outlets with opposite political stances. The boldface $[$\textbf{tokens}$]$ in brackets indicate writing bias in  \textit{Exaggeration}.
   Text in content bias is marked with frames: blue  indicates \textit{Omission} and yellow \textit{Framing}.
   Arrows point to complementary voices over the same topic and their sources are \underline{underlined} for easy reading. 
   (The original text in Chinese is shown in \ref{tb:appendix_intro}.)
     }
     % \vspace{-.3cm}
    \label{fig:intro}
\end{figure}

% \begin{figure}[htp]
%     \centering
%     \includegraphics[width=0.45\textwidth]{pictures/introfig.pdf}
%     % \vspace{-0.8em}
%     \caption{
%     Two COVID-19 news excerpts 
%     %(selected parts) 
%     % from media \textcolor{blue}{outlets}
%     from media outlets
%     % sources
%     with opposite political stances. The boldface $[$\textbf{tokens}$]$ in 
%     %but both excerpt about COVID-19. 
%     %All the 
%      %are marked by 
%      brackets indicate writing bias in  \textit{Exaggeration}.
%      %. The phrases highlighted in bold are with writing bias \textit{Exaggeration}. 
%     %The blue frame with one-way arrow stand for 
%    Text in content bias is marked with frames: blue  indicates \textit{Omission} and yellow \textit{Framing / Undue Weight}.
%    Arrows point to complementary voices over the same topic and their sources are \underline{underlined} for easy reading. 
%    (The original text in Chinese is shown in \Cref{tb:appendix_intro}.)
%     %The sentence in blue frames exhibit content bias \textit{Omission}, where the one-way arrows point to the omit aspect.
%     %which means that the pointed article omit this aspect. 
%     %The yellow frame 
%     %with bidirectional arrow 
%     %marks the text with content bias \textit{Framing / Undue Weight}.
%     %The source in each part is underline for better understanding. [intro(wrap between words?)]
%      }
%      % \vspace{-.3cm}
%     \label{fig:intro}
% \end{figure}

To further illustrate how our formulation works in the Chinese perspective, even in some situations with distinct differences in political stances, Figure \ref{fig:intro} shows two article excerpts about COVID-19, one (Article 1) from United Daily News (pro-CNP) and the other (Article 2) Liberty Times (pro-DPP).\footnote{CNP (Chinese Nationalist Party) and DPP (Democratic Progressive Party) are two major parties in Taiwan.}
Despite the opposite political stance, both of them exaggerate the outbreak in Wuhan through writing style.
For content selection, they both omitted or framed some aspects in their report, whereas their opposite positions resulted in the fact that Article 1 weighs more on the government's voice compared to Article 2's focus on the individual's voice on social media.

For data collection, we gathered 300 Chinese articles reporting COVID-19 news from 5 media outlets in Hong Kong, Taiwan, and Singapore.
Following the previous practice \cite{fan2019plain}, these articles are grouped into 100 triplets, reporting the same event and from a different source. 
In-house annotators and domain experts then manually label coarse- and fine-grained media bias on the token- and sentence-level content based on our newly designed annotation system. 
Because one sentence may contain various methods,
our multi-label assumption can be more comprehensive in understanding bias, compared with single-label applied in the existing practice \cite{aksenov2021fine,lim2020annotating}.

Based on the multi-label setup, we employ popular NLP baselines to predict both sentence-level and token-level media bias and the results show that the pre-trained BERT performs the best. 
Then, the joint training is conducted to analyze the interactions between token-level and sentence-level contexts in detecting media bias, while prior studies tend to discuss them separately \cite{fan2019plain,lei2022sentence}.
It enables better results than their separate training counterparts, which shows features within the sentence play an important role in lexical bias.
As for informational bias, which can be benefited from encoding richer context (other articles from the same triplet), is more challenging and shows that is usually signaled by external context.

When conducting the qualitative analysis to examine the advances and challenges of existing NLP models in detecting bias, we observe that, due to the small scale of the data, overfitting would be a big problem when the latest and strongest models are applied to this task. 
In social science research, media bias is usually analyzed in quantitative measure \cite{covert2007measuring}, and the whole outlet is taken as a unit.
Although smaller units, such as articles or sentences \cite{gangula2019detecting,chen2020analyzing,aksenov2021fine}, are focused when exploring its automatic detection with various computational methods \cite{iyyer2014political,baumer2015testing}, the scale of the media bias dataset is still far too small compared with other machine learning tasks.
Especially when high-quality data and labels are pursued, which require solid expertise, the cost is too high to annotate a large dataset by humans.

In further discussions, we explore whether the external corpus, which is large-scale but unlabeled, can contribute to the media bias detection task through some contextual enrichment methods.
Inspired by the support of token-sentence joint training in lexical bias and the additional context in informational bias, we consider two main directions.
One is \textbf{Data-Augmented Context Enrichment (DACE)}, which increases the size of the dataset by generating more training cases based on the token and sentence features. 
Another is \textbf{Retrieval-Augmented Context Enrichment (RACE)}, which improves information retrieval methods to select valuable information from a large outside knowledge source and feeds it to the model when a case is processed.

From the additional experiments, we notice that with the enrichment of the context, whatever method is used, the results outperform the baselines, which is also verified in the English dataset BASIL \cite{fan2019plain}.
In general, RACE methods get better results.
When the context is considered during the retrieval process, the results are further improved.
However, the performance of the model would be disturbed if the time sequence is ignored in the news-related task, where the reports are time-sensitive. We would consider this in future work.
% At last, we test our task in ChatGPT and point out the direction of future work under more powerful engines.

In summary, to the best of our knowledge, in this paper we make the following contributions: \\
\emph{
\indent $\bullet$~ We design a novel annotation system and are the first to explore Chinese media bias, contributing the first media bias benchmark in a multi-label setting.\\
\indent $\bullet$~We finetune the pre-trained models by the media bias dataset and comprehensively study the difference in the detection of lexical bias and informational bias.\\
\indent $\bullet$~We are the first to propose and verify that contextual enrichment methods, such as Data Augmentation and Information Retrieval, can benefit media bias detection tasks. Explore how external unlabeled data can contribute to such practical and small-scale tasks.}\footnote{We will make our annotated corpus publicly available upon publication.}

\section{Media Bias Structure Design}
\label{sec:label_intro}
Based on the \textbf{lexical bias} and \textbf{informational bias} bias structure proposed by Fan, Lisa, et al \citep{fan2019plain}, we go one step further to craft 4 fine-grained labels in lexical bias and 3 fine-grained labels in informational bias so that it could be clearer defined and better fitted to the Chinese context, which are listed below.

\begin{table*}[htbp]
\centering
\footnotesize
\begin{tabular}{p{0.97\textwidth}}
\toprule
\textbf{Lexical Bias Annotation Examples} \\\midrule
(a) ... They are $[$crazy for$]_\textbf{Exaggeration}$ air tickets, flying around the earth, hoping to "return to $[$the safest$]_\textbf{Exaggeration}$ place" ... \\\midrule
(b) ... some people even continue to go out to have dinner with friends. The ``$[$Buddhism$]_\textbf{Stereotype}$'' is surprising ...\\\midrule
(c) ... $[$some experts doubt the effectiveness of mass quarantine, pointing out that ``the virus will actually spread''$]_\textbf{Ventriloquism}$\\\midrule
(d) ... and the situation $[$is bound to continue to deteriorate$]_\textbf{Opinion}$ ...\\
\bottomrule
\toprule
\textbf{Informational Bias Annotation Examples}
\\\midrule
\textbf{\underline{Annotation Set 1}}\\
\textit{\underline{Article 1}}
... The Chinese song made many $[$\underline{foreign netizens} sarcastically say in a ``high-level black'' way that this ``North Korean boy'' is too cute.$]$…
$[$According to \underline{``The Paper''}, the song has a detailed and in-depth description of the mobile cabin hospital, which helps patients fight the virus and relieve tension. $]_\textbf{Framing (Positive)}$ ...
\\
\specialrule{0.00001em}{1pt}{1pt}
\textit{\underline{Article 2}}
... $[$\underline{Some netizens} denounced the creator for ``entertaining disasters''$]$. $[$\underline{The Beijing News} online commentary even denounced this as ``forcibly extolling disasters''. $]_\textbf{Framing (Negative)}$  ...
$[$\underline{Jiang Junrong} said ``A lot of places are `cleared' now. Don't we need to show a little bit of optimism? Don't we need to show a better mentality?'' $]_\textbf{Omission of Article 1}$ ...
\\
\specialrule{0.00001em}{1pt}{1pt}
\textit{\underline{Article 3}}
... $[$\underline{Some netizens} even described the video as ``seemingly back in the 1980s and 1990s''.$]$ ...
$[$Composer \underline{Jiang Junrong} said in response that ``They don't know my state of life, and I don't make any calculations''.$]_\textbf{Omission of Article 1}$
$[$\underline{The Beijing News} commented on this, saying that turning it into the material of cheerful children's songs not only dilutes the disaster background, but also eliminates the seriousness of the fight against the epidemic. $]_\textbf{Framing (Negative)}$ 
\\\midrule
\textbf{\underline{Annotation Set 2}}\\
\textit{\underline{Article 1}}
... ``Shincheonji'' chairman will hold a press conference to publicly respond to the epidemic for the first time.
\\
\textit{\underline{Article 2}}
South Korean leader of ``Shincheonji'' kneels to apologize and thank the government for fighting the epidemic.
\\
\textit{\underline{Article 3}}
Weird! The leader of ``Shincheonji'' kneels and $[$wears Park Geun-Hye's watch$]_\textbf{Sensationalist}$.
\\
\bottomrule
\end{tabular}
\caption{Our media bias annotation examples.
For simplicity, we do not show the whole annotation set for lexical bias examples. The informational bias annotations are also omitted for lexical bias examples (so as informational bias examples).
We underline the source in each part of the articles for easier understanding. (The original text refers to 
\cref{tb:appendix_example}.)}
\label{tb:example}
\end{table*}

\subsection{Lexical Bias}
Although we note there are some prior studies for lexical bias \cite{price2005framing,gentzkow2010drives,schuldt2011global}, to the best of our knowledge, there does not exist a systematic study summarizing all sub-types of lexical bias for media bias.
Building upon existing literature and also in consultation with domain experts, we design the following sub-types to measure the bias in writing styles.

\newparagraph{Exaggeration.} 
It is defined as using eye-catching wordings with strong emotional implications or describing things in a way that is more than it really is, to influence an audience \cite{schopenhauer2017delphi}. 
For example, articles in \Cref{fig:intro} use tokens ``explode'' and ``despairingly'' (in the 1st paragraph) to exaggerate the phenomenon, while other media (not in \Cref{fig:intro}) only uses ``residents fever''.

\newparagraph{Stereotype.}
Sometimes a media source may favor or attack a particular group, or a person, and describe them with wordings reflecting their stereotype to them.
We refer to this kind of bias as \textit{Stereotype}, or we can call it labeling.
It can also describe the events or people using words that may contain judgments \cite{hamborg2020media, maass1999linguistic}, e.g., \Cref{tb:example}(b) employs ``Buddhism'' to criticize the slack attitude to epidemic prevention.

\newparagraph{Ventriloquism.}
We annotate text spans as \textit{Ventriloquism} when experts or witnesses are quoted in a way that intentionally voices the author's own opinion \cite{hall201629}.
For instance, in \Cref{tb:example}(c), part of a sentence from the expert is quoted to emphasize the uselessness of mass quarantine, which may be just the author's opinion.

\newparagraph{Opinion.}
Inspired by Anand \cite{anand2007information}, this sub-type covers cases like stories that focus not on what has occurred, but primarily on what might occur, or it is just the opinion of the author.
We can also name it Speculative Content.
In \Cref{tb:example}(d), ``continue to deteriorate'' is just the speculation of the author but not the fact.

\subsection{Informational Bias}

Informational bias, a.k.a., content bias in some studies, refers to the bias reflected by selecting or emphasizing part of the event. 
It can be indicated by comparing the report with the counterpart views from other media outlets \cite{fan2019plain, van2020context, chen2020detecting}.
Our design of informational bias sub-types is mostly inspired by Baker \cite{baker1994identify}.

For certain informational bias sub-types, such as \textit{Omission}, we rely on the concept of ``stakeholders'' to help identify the informational bias more objectively: the information from different sources is first sorted out into several stakeholders, and we then compare these stakeholders within or across articles to evaluate the existence of informational bias.

\newparagraph{Omission.}
We use \textit{Omission} to describe situations where some reports ignore information that tends to disprove their claims \cite[Chapter~2]{baker1994identify}: if an article does not contain the content about stakeholders that appeared in other articles, then it has the informational bias of \textit{Omission}.
In our displayed cases, the missing descriptions of the composer's motivation for writing the song in Annotation Set 1 in \Cref{tb:example} and the citizen's situation in \Cref{fig:intro} are both omissions.

\newparagraph{Framing.}
We say an article has the informational bias of \textit{Framing} when multiple stakeholders are contained but not balanced or undue weight.
Such imbalances can be raised from the following aspects: different positions \cite[Chapter~4]{baker1994identify}, different lengths of the description \cite[Chapter~6]{baker1994identify}, and different sentiment polarities of selected content. 
The case in \Cref{fig:intro} reflects the bias on the different lengths of descriptions of the official response, while the Annotation Set 1 in \Cref{tb:example} reflects the bias on the positive and negative sentiment of the reports.

\newparagraph{Sensationalist.} 
This informational bias sub-type refers to the event that is not closely relevant to the theme, but selected to excite the greatest number of readers \cite{tannenbaum1960sensationalism}.
For example, in the Annotation Set 2 of \Cref{tb:example}, the author focuses on the same style of watch in the report about the epidemic to draw readers' attention.

\section{Dataset Construction}
\label{sec:data}

\subsection{Data Collection}

Different media outlets normally have different descriptions for the same event (as presented in \Cref{fig:intro}), providing us with an opportunity to study where bias
raises.
Motivated by this observation, in our paper, we mainly focus on studying bias from Chinese news related to the COVID-19 pandemic.
We first collect a large collection of news articles from five media outlets over three different areas or countries, ranging from January 2020 to August 2020.\footnote{Media outlets include Wen Wei Po, Apple Daily, United Daily News, Liberty Times, and Lianhe Zaobao.}
% Areas or countries include Hong Kong, Taiwan, and Singapore.}
We then filter out articles that contain only objective information, such as COVID-19 case updates.
Finally, 300 articles are selected following Fan, Lisa, et al \cite{fan2019plain} to construct 100 sets, with each set containing 3 articles from different outlets but discussing the same event.

\subsection{Human Annotation}
\label{subsec:annotation}

\begin{figure}[!t]
    \centering
    \includegraphics[width=0.45\textwidth]{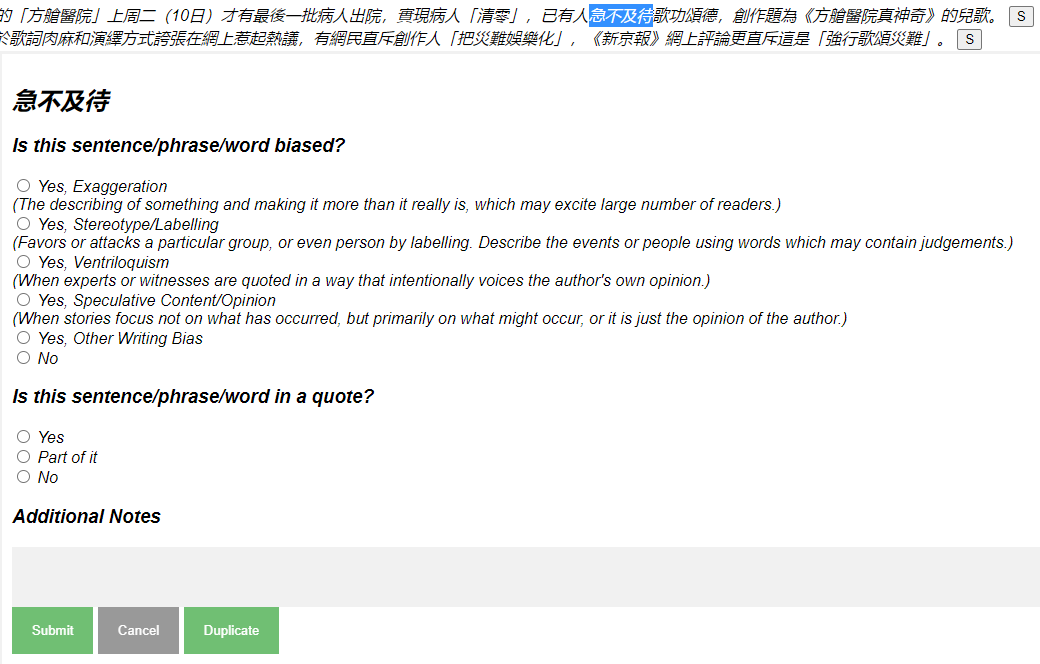}
    \caption{The interface of lexical bias annotation tool. Except for the sentence-level annotation, the annotators can also annotate in spans for token-level annotation.}
    % \vspace{-.6cm}
    \label{fig:tool_writing}
\end{figure}

\newparagraph{Annotation Process.}
Given the different nature of lexical and informational bias, we annotate them separately, though the basic tool functions are shared.
To illustrate our annotation tool, \Cref{fig:tool_writing} shows a sample interface for lexical bias while that for informational bias is put in \Cref{fig:tool_content}.
We recruit in-house annotators from a journalism background and provide them with hands-on and specific training. 
All types of bias are explained through cases and they are required to exercise, discuss, and analyze the results to warm up the annotation. 
To further mitigate the bias of annotators,  we detail our regulations to assist guideline understanding; e.g., four main source categories are exampled (officers, citizens, experts, and media) to help annotators learn stakeholders' concepts for better labeling content bias.
During the annotation, the constructed sets of articles through a web page are presented to annotators. 
Annotators should first go through all three articles in the set, and determine if a given sentence contains a certain sub-type of the bias.\footnote{To exhaustively measure the media bias from language, text spans that are in quotation marks are also considered.}
After that, media bias labels are assigned to two-level contexts, sentence-level, and token-level. 
In this way, our annotation not only indicates what media bias types are observed, but also how it is presented by certain text spans.

To further present labeling details, all the sub-types of lexical bias and informational bias that annotators can choose from (also referred to as \textbf{fine-grained} labels) are summarized in \Cref{tb:fine_grained_data_stats}.
Based on the existence of a certain bias sub-type, we also give each sentence a \textbf{coarse-grained} label, namely \textit{containing a lexical bias}, \textit{containing an informational bias}, and \textit{no bias exists}.
Both the coarse-grained and fine-grained annotations will be explored in our experiments in \Cref{sec:exp_setup}.

\newparagraph{Inter-Annotator Agreement.}
Each set is annotated by at least two in-house annotators. Cohen's Kappa \cite{10.1162/coli.07-034-R2} and F1 score is used following Fan, Lisa, et al \cite{fan2019plain} and Toprak et al \cite{toprak2010sentence} to measure the reliability between annotators in the sentence- and token-level.

The calculated Kappa value for coarse-grained annotation is 0.59; while for collected annotations with fine-grained labels, we observe 0.50 in lexical bias and 0.56 in informational bias.
The cross-annotator F1 score of coarse-grained token-level annotation is 0.76, with 0.62 in lexical bias and 0.75 in informational bias.
These are above the bar of good agreement indication~\cite{fleiss2013statistical}, while also implying the challenges in media bias annotation, especially in finer granularity.
It is because of the subjective nature of media bias and consistent with the previous findings that high inter-annotator agreements are not usually observed in subjective and fine-grained NLP annotations \cite{spinde2021neural}.
Besides, fine-grained lexical bias shows relatively lower agreement than informational attributed to more categories involved.

To further control the quality of our annotations, a domain expert is invited to manually check the in-house annotations and decide the final labels (as the gold-stand), compared to which our in-house annotators exhibit 0.89 F1 on average.
It implies acceptable annotation quality with hands-on training and in-house verification,
consistent with previous practice to use expert annotators \cite{spinde2021neural} to improve the quality and the low inter-annotator agreement for similar media bias annotation tasks through crowdsourcing \cite{lim2020annotating}.

\begin{table}
\centering
\small
\resizebox{0.45\textwidth}{!}{
\begin{tabular}{l|l|r}
\toprule
\textbf{Type} & \multirow{2}{*}{\textbf{Sub-Type (Fine-grained)}} & \multirow{2}{*}{\textbf{\# Sent.}} \\
\textbf{(Coarse)} & & \\\midrule
\multirow{5}{*}{\makecell[l]{Contain\\Lexical\\Bias}} & Exaggeration & 681 \\
 & Stereotype & 690 \\
 & Ventriloquism & 54 \\
 & Opinion & 1,318 \\\cmidrule{2-3}
 & Sent. with $\geq$1 lexical sub-type & 2,022 \\
\midrule
\multirow{4}{*}{\makecell[l]{Contain\\Informational\\Bias}} & Omission & 1,569 \\
 & Framing & 427 \\
 & Sensationalist & 67 \\\cmidrule{2-3}
 &  Sent. with $\geq$1 informational sub-type & 1,819 \\
\bottomrule
\end{tabular}
}
\caption{Breakdown statistics of sentences for different bias sub-types. 
Each sentence has a coarse-grained label (see \Cref{subsec:annotation}) and associated fine-grained bias sub-types if applicable.}
\label{tb:fine_grained_data_stats}
\end{table}

\subsection{Data Analysis}
\label{subsec:data_analysis}

\newparagraph{Number of Bias Types per Sentence.}
In total, 5,012 sentences are used for our annotation task. Among these, 59.8\% of sentences contain at least one bias sub-type label. 
Sentences in our dataset may have more than one annotated span, with each sentence or even each span possibly having different bias sub-types: we observe that 1,313 sentences (26.2\%) are such cases.
It thus inspires us to perform experiments in a multi-label framework.
The breakdown statistics for each bias sub-type are reported in \Cref{tb:fine_grained_data_stats}.

\newparagraph{Data Imbalance.}
From \Cref{tb:fine_grained_data_stats}, we can observe that the numbers of sentences in each sub-type are quite imbalanced and there is even a more than 20 times difference.
One of the reasons is that some sub-types, for example, \textit{Sensationalist} seldom appear in health news, though widely applied in other domains, e.g., entertainment.
The data imbalance issue may concretely challenge models to characterize certain media bias sub-types, although data augmentation may potentially mitigate the problem, which will be discussed later.

\newparagraph{Sentence Length.}
We calculate the average length for sentences containing bias sub-types. 
We observe that the sentences with informational bias on average contain 52.2 tokens, which are longer compared to those with lexical bias (19.2 tokens).
One possible explanation is that the informational bias needs more length and context to present compared with lexical bias.

\section{Chinese Media Bias Detection}
\label{sec:exp_setup}

% We first conduct some basic experiments on our dataset and analyze the results to test whether our dataset is feasible for the media bias detection task.

\subsection{Basic Task Design}
\label{subsec:main_exp_settings}

\newparagraph{Sentence-level Classification.}
We first study whether it is feasible to develop automatic methods to measure the media bias from the text.
To this end, we design a sentence-level multi-label bias classification task.
Specifically, we consider the following settings (listed as four columns in \Cref{tb:multilabel_result}). 
In \textsc{Coarse-Grained} classification, we only consider whether a sentence has any type of lexical bias or informational bias, using coarse-grained labels discussed in \Cref{subsec:annotation}.
While in fine-grained classification, the model needs to specify 4 sub-types of lexical bias (denoted as \textsc{Fine-Grained Lexical}) or 3 sub-types of informational bias (henceforth \textsc{Fine-Grained Informational}). 
To further test the model capability, we combine all bias sub-types as a 7-label multi-label classification task (named as \textsc{Fine-Grained Combined}).

\newparagraph{Token-level Sequence Tagging.}
Treating each sentence as a whole and predicting media bias labels is clearly insufficient for studying bias -- we are more interested in identifying tokens that cause certain types of bias. 
Our annotated dataset supports this kind of experiment, as annotators are asked to select text spans that trigger bias during annotation (discussed in \Cref{subsec:annotation}).
We thus formulate our token-level media bias detection as a sequence tagging task, with the goal of marking the bias span tokens in each sentence.
All annotated spans are first transformed into BIO scheme (with B tag denoting \textit{begin}, I for \textit{intermediate}, E for \textit{end}, S for \textit{single character} and O for \textit{outside}) and then used for model training.

\newparagraph{Joint Training.}
Then, we detect the bias by joint training of sentence- and token-level classification tasks to examine the interactions of the whole sentence and local tokens. 
In the warm-up period, two classifiers are trained iteratively: after 3 epochs of training of sentence-level classification, the parameters in the classifier layer for the sentence are frozen, and then 3 epochs of training are conducted in the token-level classifier. 
The losses of two classification tasks are added together as the optimization target after the warm-up steps.
% }

\begin{figure}[!t]
    \centering
    \includegraphics[width=0.45\textwidth]{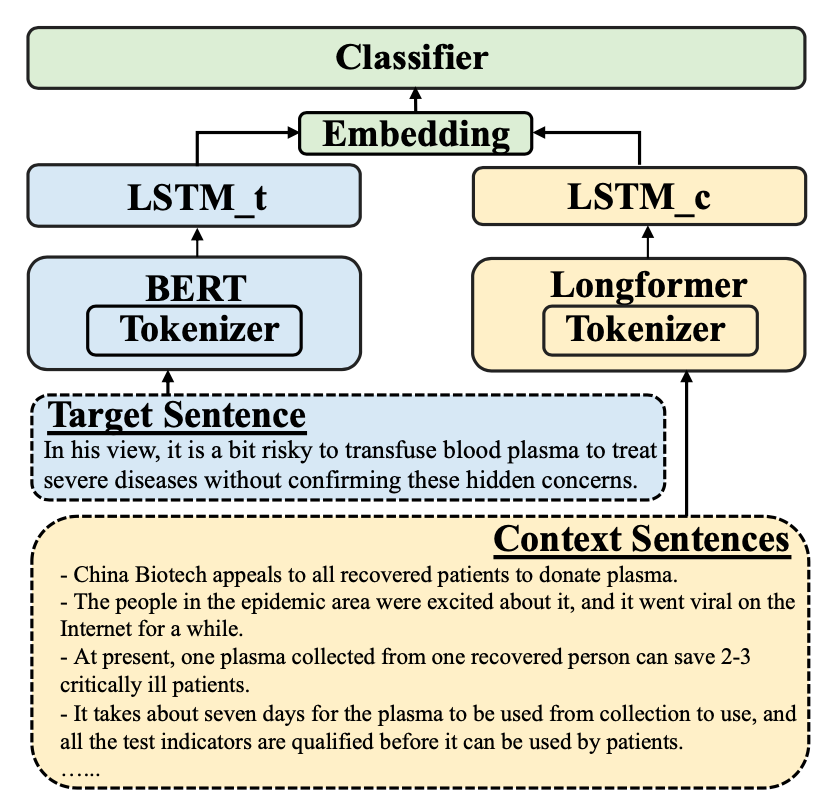}
    \caption{The structure of our model with context feature added. The target sentence is encoded by BERT and an LSTM layer, context sentences are encoded by Longformer and an LSTM layer. Their embeddings are concatenated as the input of the classifier for final results. In this case, the context sentences are all confident with the plasma treating method. They will be helpful for judging the target sentence, which doubts the feasibility of this method.
    % \red{It would be better to redesign this graph.}
    }
    \label{fig:context_model}
\end{figure}

\newparagraph{Context Feature Adding.}
\label{subsec:context_adding}
Based on the human annotation process and previous work \cite{van2020context}, we also want to learn the effects of additional context, and the experiment is conducted by adding sentences in the same sets as the context feature.
As shown in \Cref{fig:context_model}, we first encode the target sentence through BERT and at the same time, summarize the context sentences, which are from the same set as the target sentence. 
Due to the length limitation of BERT, we use the Longformer pre-train model to process the context content.
In the next stage, after encoding by two separate LSTM modules, the representation of the target sentence is concatenated with the context sentence representation and then classified to obtain sentence-level predictions.

% \newparagraph{Large Language Model (GPT).}
% Recently, ChatGPT, a strong model that interacts in a conversational way, was released by OpenAI. 
% It performs well in almost every task that machines have attempted to do in human current expectations.
% We also test our task on ChatGPT to find its strengths and weaknesses, so that we can better plan our future work and cooperate with those strong large language models.

\begin{table*}
\centering
\resizebox{0.98\textwidth}{!}{
\begin{tabular}{c|ccc|ccc|ccc|ccc}
\toprule
 \multirow{2}{*}{\textbf{Method}} & \multicolumn{3}{c|}{\textbf{Coarse-Grained}}   
 & \multicolumn{3}{c|}{\textbf{Fine-Grained Combined}} 
 & \multicolumn{3}{c|}{\textbf{Fine-Grained Lexical}} 
 & \multicolumn{3}{c}{\textbf{Fine-Grained Informational}} \\\cmidrule{2-13}

 & \textbf{HL} ($\downarrow$) & \textbf{JI} ($\uparrow$) & \textbf{F1} ($\uparrow$)
 & \textbf{HL} ($\downarrow$) & \textbf{JI} ($\uparrow$) & \textbf{F1} ($\uparrow$)
 & \textbf{HL} ($\downarrow$) & \textbf{JI} ($\uparrow$) & \textbf{F1} ($\uparrow$)
 & \textbf{HL} ($\downarrow$) & \textbf{JI} ($\uparrow$) & \textbf{F1} ($\uparrow$) \\\midrule
 
 Random & 0.484 & 0.247 & 0.396 & 
 0.497 & 0.110 & 0.199 &
 0.497 & 0.109 & 0.197 & 
 0.496 & 0.106 & 0.191 
 
\\\midrule

GPT3.5 (Zero-Shot) & 
0.328 & 0.045 & 0.085 & 
  0.135 & 0.046 & 0.087 & 
  0.131 & 0.120 & 0.214 & 
  0.162 & 0.036 & 0.070
  \\
GPT3.5 (Few-Shots) &  
0.334 & 0.123 & 0.220 & 
  0.128 & 0.096 & 0.175 & 
  0.136 & 0.140 & 0.245 & 
  0.150 & 0.072 & 0.134

 \\\midrule\midrule
 
 SVM & 0.450 & 0.200 & 0.333 & 
0.355 & 0.122 & 0.218 & 
0.393 & 0.093 & 0.170 & 
0.303 & 0.168 & 0.288\\
 
 CNN & 0.299 & 0.360 & 0.529 &
0.116 & 0.200 & 0.333 & 
0.107 & 0.314 & 0.478 & 
0.142 & 0.232 & 0.377 \\
 
 RNN & 0.287 & 0.195 & 0.326 & 
0.124 & 0.135 & 0.238 & 
0.124 & 0.203 & 0.338 & 
\underline{\textbf{0.126}} & 0.189 & 0.317\\
 
 LSTM & 0.263 & 0.325 & 0.491 & 
0.124 & 0.252 & 0.402 & 
0.113 & 0.289 & 0.449 & 
0.136 & 0.231 & 0.375\\
 
 BERT & \underline{\textbf{0.259}} & \underline{0.381} & \underline{0.552} & 
\underline{0.115} & \underline{0.342} & \underline{0.510}  & 
\underline{0.101} & \underline{0.390} & \underline{0.561} &
0.129 & \underline{0.310} & \underline{0.474}\\

 \midrule\midrule
BERT+context & 0.271 & \textbf{0.396} & \textbf{0.568} & 
\textbf{0.112} & \textbf{0.352} & \textbf{0.521} & 
0.105 & 0.375 & 0.545 & 
0.134 & \textbf{0.321} & \textbf{0.486} 
\\\midrule
Joint Train & 0.260 & 0.324 & 0.490 & 
0.119 & 0.300 & 0.461 & 
\textbf{0.095} & \textbf{0.421} & \textbf{0.592} & 
\textbf{0.126} & 0.294 & 0.454

\\
\bottomrule
\end{tabular}
}
\caption{Results for sentence-level classification, measured by Hamming loss (HL), Jaccard index (JI), and weighted F1 metrics.
The best performance for models without additional information 
% (thus last two rows are excluded \shi{not last two rows anymore}) 
is underlined. The best performance of all is marked bold.
$\uparrow$ indicates that a higher value is better and $\downarrow$ is the opposite, and the scale of HL, JI, and F1 are from 0 to 1.
% \shi{state the scale of HL and JI, i.e. from 0 to 1?}
The design of tasks is discussed in \Cref{subsec:main_exp_settings}.
}
\label{tb:multilabel_result}
\end{table*}

\subsection{Experimental Setup}
\label{subsec:setting1}

\newparagraph{Models.} Our non-neural models include the Support Vector Machine (SVM) classifier in sentence-level classification and the Conditional Random Field (CRF) in token-level sequence tagging.
Unigram and bigram features are used in SVM, while we only apply unigram features to CRF. Also in SVM, One-vs-the-rest (OvR) multi-class strategy is applied. 

We train our dataset using a set of neural models, including CNN \cite{lecun1998gradient}, RNN \cite{elman1990finding} and LSTM \cite{hochreiter1997long} models, 
300-dimension pre-trained Chinese embedding provided by Li et al \cite{li2018analogical} is applied in these neural models.
We also experiment with pre-trained models including BERT \cite{devlin2018bert} with pre-trained tokenizer and pre-trained model, 
which is conducted through Hugging Face \cite{wolf2020transformers} and \textit{Chinese} version is chosen\footnote{\textit{BertForSequenceClassification} is used for sentence-level classification and \textit{BertForTokenClassification} is used for token-level classification.}.
Then we fine-tune BERT on our dataset by AdamW \cite{loshchilov2017decoupled} optimizer.
When the Longformer \cite{beltagy2020longformer} model is applied, the basic setting is the same.

% \newparagraph{Large Language Model (GPT).}
Recently, ChatGPT, a strong model that interacts in a conversational way, was released by OpenAI. 
It performs well in almost every task that machines have attempted to do in human current expectations.
We also test our task on ChatGPT to find its strengths and weaknesses, so that we can better plan our future work and cooperate with those strong large language models.
Specifically, we use the model \textit{gpt-3.5-turbo-16k} through LangChain and give the descriptions of each type of bias. The prompts are shown in \ref{app:prompts}.

In all the experiments, we include a random baseline model, which randomly selects half of the sentences as containing a specific bias sub-type.

\newparagraph{Data Splits and Evaluation Metrics.}
We randomly divide our annotated 100 sets by 8:1:1 into training, validation, and test splits, with each containing 3,915 sentences, 608 sentences, and 489 sentences. We then select the model with the lowest validation loss used for evaluation. 
In our experiments, the learning rate is 5e-6, the batch size is 16, and the maximum length for each sentence is 128. In the context feature adding task, we randomly selected 8 sentences from the set instead of the whole set due to the limitation of our computing sources.
For GPT-related experiments, in the few-shot setting, we randomly sample 16 cases from the training set, and evaluations are done on the test set.

We use Hamming loss, Jaccard index, and F1 metrics to measure our model performance. The micro average method is used for the Jaccard index and F1 score in sentence-level classification tasks.

% In GPT test experiments, we use the model \textit{gpt-3.5-turbo-16k} and give the descriptions of each type of bias. The prompts are shown in \ref{app:prompts}.
% When we conduct the few-shots experiments, we randomly sample 16 cases from the training set. 
% \shi{GPT related experimental settings should be incorporated into a paragraph above}

\subsection{Basic Results}
\label{sec:exp_results}

\subsubsection{Sentence-level Results}
\label{sec:sent_level}

We start the experimental discussions with the comparison results of sentence-level classification in \Cref{tb:multilabel_result} and the following observations can be drawn.

First, neural models in general outperform their non-neural counterparts.
It indicates our task is challenging, where $n$-gram features may fail to signal media bias.
Among neural models, BERT performs the best, especially in fine-grained scenarios, possibly due to its superior semantic understanding ability gained via large-scale pre-training and more complex structure.

Second, within fine-grained predictions, most models perform worse for informational bias than lexical.
The possible reason is informational bias may require more knowledge to understand while some token-level indicators may sufficiently signal lexical bias.
To probe into this point, we feed the BERT model with richer context (content from other articles in the same set), enabling it to compare different source's information.

The results of this BERT+context variant (BERT with context feature adding) are also listed in Table \ref{tb:multilabel_result}.  
It is observed that the learning of fine-grained content bias can indeed benefit from richer context, because models can compare the content selected by different authors and better capture semantics. 
This finding is consistent with that drawn by Fan et al \cite{fan2019plain} for coarse-grained informational bias (as well as our coarse-grained results), and together inspire the future potential in exploring automatic context formation and learning to understand informational bias, which also discuss in \Cref{sec:irmethods}.

\subsubsection{Token-level Results}
\label{sec:token_level}

Then, we examine how models perform to understand token-level bias and report the results in \Cref{tb:token_result}. 
We observe that both non-neural and neural models present fine recall, whereas fine-tuned BERT still outperforms other baselines, indicating the helpfulness of language pre-training and complex structure for token-level bias understanding as well.

We also find the overall results for lexical bias are much better than informational bias.
It is probably because informational bias tends to be reflected with richer context, rendering longer spans marked as its indicators (as discussed in \Cref{subsec:data_analysis}) and lower chances to perfectly predict for all involved tokens.

\begin{table*}
\centering
\small
\resizebox{0.85\textwidth}{!}{
\begin{tabular}{c|ccc|ccc|ccc}
\toprule
\multirow{2}{*}{\textbf{Method}} & \multicolumn{3}{c|}{\textbf{Lexical}} 
 & \multicolumn{3}{c|}{\textbf{Informational}}
 & \multicolumn{3}{c}{\textbf{Combined}}  \\\cmidrule{2-10}
 
  & \textbf{Precision} & \textbf{Recall} & \textbf{F1}
 & \textbf{Precision} & \textbf{Recall} & \textbf{F1}
& \textbf{Precision} & \textbf{Recall} & \textbf{F1}\\\midrule

Random & 0.745 & 0.201 & 0.298 & 
0.500 & 0.197 & 0.279 & 
0.478 & 0.199 & 0.277 \\\midrule

CRF & 0.728 & 0.853 & 0.786 & 
0.434 & 0.529 & 0.440 & 
0.432 & 0.452 & 0.408 \\

RNN & 0.815 & 0.847 & 0.825 & 
0.591 & 0.602 & 0.589 & 
0.609 & 0.605 & 0.600 \\

LSTM & 0.810 & 0.851 & 0.821 & 
0.598 & 0.609 & 0.594 & 
0.610 & 0.603 & 0.594 \\

BERT & \underline{\textbf{0.860}} & \underline{0.860} & \underline{0.858} & 
\underline{0.649} & \underline{0.651} & \underline{0.650} & 
\underline{\textbf{0.675}} & \underline{\textbf{0.675}} & \underline{\textbf{0.674}} \\\midrule\midrule

Joint Train & 0.858 & \textbf{0.866} & \textbf{0.861} & 
\textbf{0.673} & \textbf{0.658} & \textbf{0.661} & 
0.674 & 0.673 & 0.671 \\

\bottomrule
\end{tabular}
}
\caption{Experimental results on token-level bias classification task. We use the same notations as in \Cref{tb:multilabel_result}.}
\label{tb:token_result}
\end{table*}

\subsubsection{Joint Training Results}
\label{sec:joint_train}

We have discussed sentence-level and token-level results to learn that lexical bias and information bias get effects from different levels of context.
To further examine their interactions, we jointly train a classifier to predict two-level media bias with multi-task learning (henceforth Joint Train). 
The results are reported separately in the last row in \Cref{tb:multilabel_result,tb:token_result} for easier comparison purposes with other methods.

For sentence-level results, Joint Train provides more performance gain to fine-grained lexical bias detection, while informational bias benefits more from additional context (discussed in \Cref{sec:sent_level}).
It is because, as we speculated, the judgment basis of lexical and informational bias is quite different; the former may require more knowledge and the latter may be explicitly indicated by a few tokens.
As a result, lexical bias prefers features that reflect internal semantics, such as tokens, while adding external information would better benefit informational bias learning.  

On the token-level, the benefit of joint training is observed universally for both lexical and informational bias, suggesting a sentence-level view gained on comprehensive semantics helps capture token-level bias.

\subsection{Analysis}

\subsubsection{Analysis of Fine-grained Classification}
\label{subsec:fine-grained-analysis}

The above discussions mostly concern models' averaged fine-grained results. 
Here we further probe into each media bias sub-type and find model performs differently over varying fine-grained labels.
A factor analysis is hence conducted over the output of Joint Train and BERT+context, respectively champions the overall lexical and informational bias detection (Table \ref{tb:multilabel_result}). 
The following findings are drawn.

First, models' performances are largely affected by the available training data scales, whereas label imbalance obviously exhibits in our fine-grained annotations (Table \ref{tb:fine_grained_data_stats}).
Take informational bias annotation as an example: roughly 75\% of the sentences are labeled as \textit{Omission}.
Bert+Context thus achieves a 0.52 F1 in detecting \textit{Omission}, much higher than the other two sub-types. 
These shed light on the importance of advancing robust training with imbalanced labels for fine-grained media bias prediction. 

Second, the nature of labels also plays a crucial role.
For lexical bias sub-types, Joint Train better predicts \textit{Exaggeration} and \textit{Stereotype} (both exhibiting 0.65 F1) compared to a 0.55 for \textit{Opinion}, although the latter almost has twice training samples.
It might be because most cases of the former two sub-types are explicitly indicated by certain tokens like ``crazy'' (Table \ref{tb:example}(a)), ``explode'' (Figure \ref{fig:intro}), whereas expressions in \textit{Opinion} are relatively more implicit and uncertain, sometimes requiring higher-level knowledge for understanding.
It calls for future attention to injecting knowledge to understand media bias in implicit languages. 

\subsubsection{Analysis of Large Language Model}

From \Cref{tb:multilabel_result}, we observe that ChatGPT is strenuous in media bias detection. 
Although it is improved by few-shots experiments, the performance is still not as good as the fine-tuned model.
To further explore the reason for poor performance, we experiment with the detection of only lexical bias and informational bias. 
We observe it has high precision but low recall scores. For example, in lexical bias, it got 0.846 in precision, but only 0.065 in the recall, which means the model is not sensitive to bias and misses most of the biased sentences. 

We can also notice that there is a big difference in the performance of ChatGPT between Fine-Grained Lexical and Fine-Grained Informational bias detection tasks.
It is harder for ChatGPT to tell informational bias.
During our testing, we noticed that sometimes ChatGPT would request more context information instead of returning the answer directly.
However, when giving the context, the limitation of input length also exists. Also, the model will omit some information when the text is too long.
It motivates us to do further research in context enrichment to supply valuable information.

% But, when a single sentence or a single article is given, usually requests more information before giving the answer, such as the source of the report, or related news, which is similar to the contexts in our experiment setting.
% However, the limitation in length also exists, the model will omit some information when the text is too long.
% So, it is quite helpful to do further research in context enrichment to supply valuable information.

% Although GPT is a strong engine, that is good at some basic language tasks, such as translation and summarization. It is also strenuous in media bias detection.

\section{Context Enrichment Methods and Experiments}
Inspired by the insights in our basic experiments \Cref{sec:exp_results}, we explore some methods to enrich the context for our model to better understand bias.
We consider two main directions of context enrichment, one is Data-Augmented Context Enrichment (DACE), which increases the size of the training set and another is Retrieval-Augmented Context Enrichment (RACE), which can supply more knowledge when training.
In order to test the generality of our methods, we also conduct experiments on the English dataset BASIL \cite{fan2019plain}.
In this section, we will introduce how we improve these methods for media bias detection tasks.

\subsection{Data-Augmented Context Enrichment}
We fine-tune the BERT with BASIL and our dataset as the baselines of the following several methods.
In the selection of data augmentation methods, we choose some traditional supervised learning methods\cite{xie2020unsupervised} that augmented the sentence at the token level. 
We also attempt a semi-supervised learning method that augments the data at the sentence level.

\newparagraph{Synonym Replacement (SR)}
Randomly choose one word that is not a stop word and replace the word with its synonym. 
This operation is repeated $n$ times.

\newparagraph{Random Insertion (RI)}
Randomly choose one word that is not a stop word in the sentence, and find its synonym. The synonym word will be inserted into a random position in the sentence.
This operation is repeated $n$ times.

\newparagraph{Random Swap (RS)}
Randomly choose two words in the sentence and swap their position.
This operation is repeated $n$ times.

\newparagraph{Random Deletion (RD)}
Remove each word in the sentence with a random possibility $p$.

\newparagraph{Easy Data Augmentation (EDA)}
EDA consists of the four methods we mentioned above, SR, RI, RS, and RD.
The policies are selected according to the randomly produced possibilities.

\label{sec:data-aug}
\newparagraph{Unsupervised Data Augmentation (UDA)}
Recently, it is common to apply the concept of consistency training to semi-supervised learning,
which is to say, the model predictions should be invariant between the original unlabeled data and the unlabeled data with noise input.
One of the important works is UDA(Unsupervised Data Augmentation) \cite{xie2020unsupervised}, which adds noise by augmentation methods, such as back translation, TF-IDF word replacement, and so on.
To align with our other data-augmented context enrichment methods, EDA is applied as the noise-adding module.

The loss function is constructed by two parts and balanced by a parameter $\lambda$, which is usually set as $1$ \cite{xie2020unsupervised}. 
One part is supervised cross-entropy loss of the labeled data (denoted as $L_{sup}$) and another is unsupervised consistency training loss, denoted as $L_{unsup}$, which is used to minimize a divergence metric between the two distributions, original data $P_{ori}$, and the data after augmented $Q_{aug}$. The equations are listed below:

\begin{equation}
    L_{unsup} = D_{KL}(P_{ori}||Q_{aug})
\end{equation}
\begin{equation}
    L = L_{sup} + \lambda L_{unsup}
\end{equation}

\subsection{Retrieval-Augmented Context Enrichment}

In this part, the result of fine-tuned BERT is also regarded as the baseline of RACE methods, without any information from outside knowledge sources.

In this task, for every sentence in the dataset (denoted as a \textbf{target sentence}), some methods are used to retrieve related sentences from knowledge sources (denoted as \textbf{context sentences}) for bias determination.
Then, the model of context feature adding in \Cref{subsec:main_exp_settings} is applied in this part. 
In our setting, the articles in the same set report the same event or focus on the same problem, which are important references for annotators to compare and determine the bias.
So we regard the result when the context sentences are from the same set as the target sentence as the ground truth of retrieval tasks.

\newparagraph{BM25}
BM25\cite{robertson2009probabilistic} is a traditional method that calculates the relevance of a query and a document based on the probabilistic retrieval framework.
For each target sentence, we regard it as a query and then calculate the BM25 scores of the sentences in knowledge sources.
The top 10 sentences are retrieved as context sentences of the target sentence according to BM25 scores.

\newparagraph{SimCSE}
SimCSE\cite{gao2021simcse} is a simple contrastive learning framework by using ''entailment'' pairs as positives and ''contradiction'' pairs as negatives to optimize the distance between sentence embedding.
The distance between the embedding of sentences with similar semantics is closer.
We fine-tune the pre-trained SimCSE model supplied by Gao et al \cite{gao2021simcse} using our datasets.
According to our setting of ground truth, the target sentence with the sentence in the same set is regarded as a positive pair, and the target sentence with a sentence that is not in the same set is regarded as a negative pair.
In the retrieval step, we calculate the Cosine similarities between the target sentence embedding and the embedding of each sentence in the knowledge sources, then the top 10 are retrieved as context sentences.

\newparagraph{Dense Passage Retrieval (DPR)}
The method DPR\cite{karpukhin2020dense} is first proposed in the open-domain question-answering task.
This model aims to rank the passage containing the answer top and put other passages in low ranking so that valuable information would be retrieved to guarantee the correct answer. 

Inspired by this concept, we design our training on the DPR model based on our datasets. 
Each instance in the training data consists of a target sentence and several context sentences, denoted as $\{(t_i,{c^+}_{i,1},...,{c^+}_{i,n},{c^-}_{i,1},...,{c^-}_{i,m})\}_{i=1}^{p}$. 
The $n$ sentences from the same set that the target sentence is in are positive sentences, and the other $m$ are negative sentences.
Following the previous work, we calculate the similarity between the sentences using the embedding output from encoders, which are denoted as $E_t$ and $E_c$.
The model is optimized by BCELoss as the equation $(4)$. 

\begin{equation}
    sim(t,c) = E_t(t)^TE_c(c)
\end{equation}

\begin{multline}
    L(t_i,{c^+}_{i,1},...,{c^+}_{i,n},{c^-}_{i,1},...,{c^-}_{i,m}) =\\
- \sum_{p=1}^{n} \log \sigma(sim(t_i,{c^+}_{i,p}))-\sum_{q=1}^{m} \log \sigma(1-sim(t_i,{c^-}_{i,q}))
\end{multline}

Finally, the top 10 sentences are retrieved as context sentences according to the similarity scores.

\newparagraph{Combining Titles.}
Sometimes, the meaning of complex information cannot be correctly reflected in a single sentence. The content of the whole article that the sentence is in is also necessary to be considered.
Usually, the title of a piece of news summarizes the main content.
So, we conduct the experiments that feed the title as the sentence's context into the model, which saves a lot of memory compared with adding the whole report.

Based on the BM25 and SimCSE model, we concatenate a sentence with its report title for training and selection.
As for DPR model, which is more flexible, we add another encoder for the title sentences. Three encoders in total are trained to get the embedding of target, context, and title sentences.
Then the embedding of each sentence is concatenated with the embedding of its article title for further similarity calculation.

\subsection{Experiments Setup}

\newparagraph{Data.}
In the experiments, we use two main datasets, one is an English dataset BASIL, and another is our Chinese dataset.
We also expand the two datasets with a large number of unlabeled reports online as the unsupervised dataset in UDA and knowledge sources in the information retrieval task.
For BASIL, we use a large dataset consisting of 754,000 news articles \cite{kiesel2019semeval}. 
For our Chinese dataset, we have about 120,000 unlabelled sentences of the reports that crawled from the COVID topic in several media outlets in the same period.

\newparagraph{Setup}
Most of our media bias detection tasks are based on BERT \cite{devlin2018bert}. 
The experiments are conducted through Hugging Face \cite{wolf2020transformers}, and the pre-trained tokenizer and model are applied, according to the language of BASIL and our dataset, which are BERT$_{base-cased}$ and BERT$_{base-chinese}$.
The synonym packages for EDA are also selected according to the languages.

Due to the data extreme imbalance in BASIL, we just simply copy the sentences in the minor class to balance the dataset, while in the data augmentation supervised learning experiments, we select sentences with different labels by the same scale from the augmented dataset. 
Also, the final result is the average of the results with 5 different random seeds.
Precision, Recall, and F1 score are selected to measure the performance in BASIL according to the previous work \cite{fan2019plain}.
Other settings are the same as the basic experiments in \Cref{subsec:setting1}.

% \subsection{Results}

% \section{Context Enrichment Results}

\begin{table*}
\centering
\small
\resizebox{0.99\textwidth}{!}{
\begin{tabular}{l|ccc|ccc|ccc|ccc|ccc|ccc}
 
\toprule
\multirow{3}{*}{\textbf{Method}} & \multicolumn{6}{c|}{\textbf{BASIL}} 
 & \multicolumn{12}{c}{\textbf{Our dataset}} \\\cmidrule{2-19}
 & \multicolumn{3}{c|}{\textbf{Lexical}} & \multicolumn{3}{c|}{\textbf{Informational}} 
 & \multicolumn{3}{c|}{\textbf{Coarse-Grained}}   
 & \multicolumn{3}{c|}{\textbf{Fine-Grained Combined}} 
 & \multicolumn{3}{c|}{\textbf{Fine-Grained Lexical}} 
 & \multicolumn{3}{c}{\textbf{Fine-Grained Informational}}  \\\cmidrule{2-19}
 & \textbf{P} ($\uparrow$) & \textbf{R} ($\uparrow$) & \textbf{F1} ($\uparrow$)
 & \textbf{P} ($\uparrow$) & \textbf{R} ($\uparrow$) & \textbf{F1} ($\uparrow$)
 & \textbf{HL} ($\downarrow$) & \textbf{JI} ($\uparrow$) & \textbf{F1} ($\uparrow$)
 & \textbf{HL} ($\downarrow$) & \textbf{JI} ($\uparrow$) & \textbf{F1} ($\uparrow$)
 & \textbf{HL} ($\downarrow$) & \textbf{JI} ($\uparrow$) & \textbf{F1} ($\uparrow$)
 & \textbf{HL} ($\downarrow$) & \textbf{JI} ($\uparrow$) & \textbf{F1} ($\uparrow$) 
 \\\midrule

 BERT & 0.103 & 0.671 & 0.178 & 0.228 & \textbf{0.729} & 0.344 
 & 0.259 & 0.381 & 0.552 & 0.115 & 0.342 & 0.510 
 & 0.101 & 0.390 & 0.561 
 & 0.129 & 0.310 & 0.474 \\
  BERT+Context & 0.129 & 0.589 & 0.209 & 0.314 & 0.624 & 0.415 & 0.271 & 0.396 & 0.568 & 0.112 & 0.352 & 0.521 & 
0.105 & 0.375 & 0.545 & 0.134 & 0.321 & 0.486
 \\
 \midrule\midrule
Synonym Replace & 0.111 & 0.627 & 0.185 & 0.283 & 0.529 & 0.342 
& 0.265 & 0.428 & 0.599 & 0.115 & 0.311 & 0.474 & 0.097 & 0.363 & 0.533 
& 0.143 & 0.329 & 0.496 \\
Random Insertion  & 0.111 & 0.635 & 0.187 & 0.273 & 0.498 & 0.322 
 & 0.260 & 0.423 & 0.594 & 0.113 & 0.338 & 0.505 & \textbf{0.094} & 0.397 & 0.568 
 & 0.142 & \textbf{0.343} & \textbf{0.511} \\
Random Swap  & 0.111 & 0.603 & 0.184 & 0.276 & 0.543 & 0.346 
& 0.270 & 0.409 & 0.580 & 0.115 & 0.332 & 0.498 & 0.097 & 0.390 & 0.561 
& 0.137 & 0.325 & 0.491 \\

Random Deletion  & 0.110 & 0.654 & 0.188 & 0.269 & 0.539 & 0.331 
& 0.267 & 0.409 & 0.580 & 0.116 & 0.302 & 0.464 & 0.100 & 0.373 & 0.543 
& 0.138 & 0.341 & 0.508 \\\midrule

EDA & 0.116 & 0.649 & 0.194 & 0.285 & 0.525 & 0.345 
& 0.288 & 0.406 & 0.578 & \textbf{0.110} & 0.349 & 0.517 & 0.098 & 0.382 & 0.552 
& 0.152 & 0.331 & 0.498\\\midrule
UDA 
& 0.133 & \textbf{0.688} & 0.222 & 0.260 & 0.651 & 0.371 
& 0.260 & \textbf{0.437} & \textbf{0.608}
& 0.118 & 0.343 & 0.511
& 0.096 & 0.408 & 0.580
& 0.156 & 0.321 & 0.486
\\\midrule\midrule

BM25  & 0.134 & 0.546 & 0.214 & 0.302 & 0.543 & 0.385 
 & 0.268 & 0.423 & 0.594 & 0.114 & 0.341 & 0.508
  & 0.102 & 0.372 & 0.543 & 0.128 & 0.311 & 0.475 \\

SimCSE  & 0.146 & 0.616 & 0.235 & 0.316 & 0.615 & 0.414 
 & 0.274 & 0.415 & 0.586 & 0.112 & 0.350 & 0.518 
  & 0.101 & \textbf{0.409} & \textbf{0.581} & \textbf{0.123} & 0.322 & 0.487 \\

DPR  & 0.139 & 0.588 & 0.223 & 0.315 & 0.621 & 0.415 
 & \textbf{0.246} & 0.430 & 0.602 & 0.116 & 0.333 & 0.499 
  & 0.101 & 0.388 & 0.559 & \textbf{0.123} & 0.314 & 0.478 \\
 \midrule

BM25+title  & 0.122 & 0.419 & 0.188 & 0.290 & 0.552 & 0.374
 & 0.260 & 0.424 & 0.596 & 0.116 & 0.328 & 0.494 
  & 0.113 & 0.363 & 0.533 & 0.131 & 0.273 & 0.429 \\
 
SimCSE+title  & 0.137 & 0.602 & 0.222 & 0.300 & 0.642 & 0.404 
 & 0.264 & 0.425 & 0.597 & 0.122 & 0.361 & 0.530 
  & 0.102 & 0.375 & 0.545 & \textbf{0.123} & 0.314 & 0.478 \\

DPR+title  & \textbf{0.147} & 0.635 & \textbf{0.239} & \textbf{0.319} & 0.616 & \textbf{0.419} 
 & \textbf{0.246} & 0.430 & 0.602 & \textbf{0.110} & \textbf{0.378} & \textbf{0.549}
 & 0.098 & 0.398 & 0.570 & \textbf{0.123} & 0.316 & 0.480 \\

\bottomrule
\end{tabular}
}
\caption{Experimental results of Data Augmentation and Information Retrieval methods on dataset BASIL and ours. We use the same notations as in \Cref{tb:multilabel_result}.}
\label{tb:combined_result}
\end{table*}

\subsection{Results of DACE}
% \subsection{Data-Augmented Enrichment}

% The EDA and UDA methods enrich the phrases and sentences in the training, we can regard them as corpus enrichment methods.
In DACE, SR, RI, RS, RD, and EDA are supervised learning methods with augmentation in the token level, while UDA is a semi-supervised learning method that augments the data in the sentence level.

\newparagraph{Token-Level Enrichment.}
From \Cref{tb:combined_result}, we observe that the policies of EDA in general enable BERT to perform better, although there are minor differences among each policy, indicating the overall helpfulness of data augmentation on our task.
All four types of policies help to increase the scale of the training set, while the policy RI enriches the word corpus without content reduction, which shows better performance than other policies.
% Compared with the policies SR and RD, RI shows better performance.

In conclusion, implementing the four policies together increases the robustness of the performance, and gained improvements in all tasks.
This further confirms that EDA is suitable to be used as a noise-adding approach in UDA, which can maintain the consistency of label distribution in media bias detection tasks.

\begin{figure}[!t]
    \centering
    \includegraphics[width=0.45\textwidth]{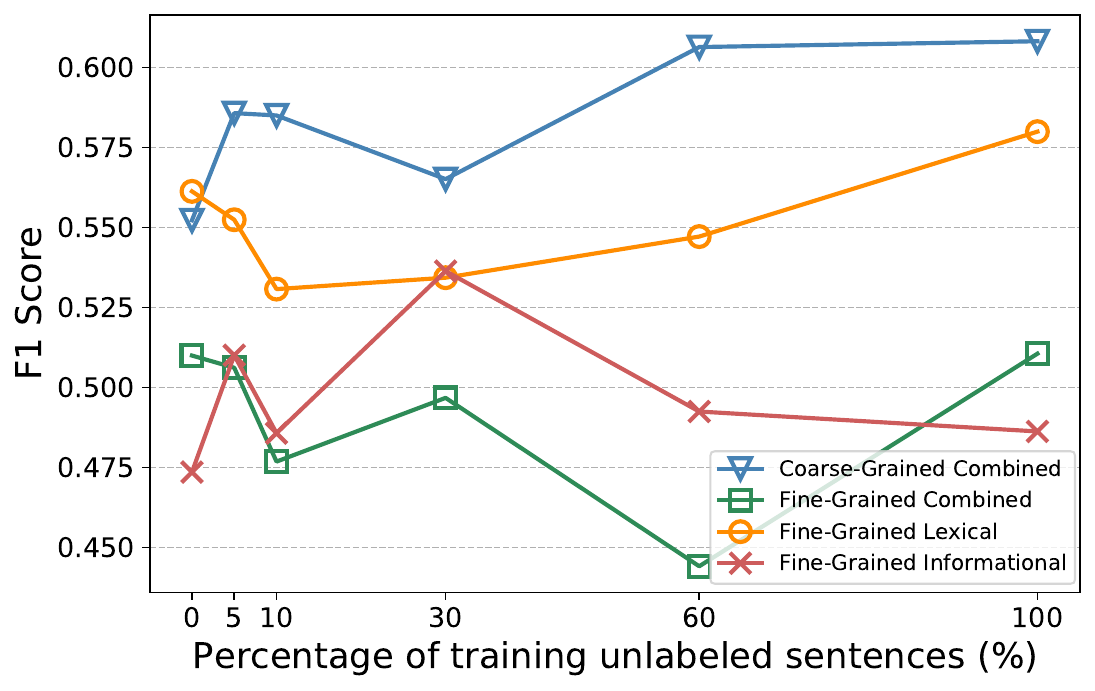}
    \caption{Results of UDA with different scales of unlabeled data.
   $x$-axis: varying percentage of the extended dataset as unlabeled sentences when training UDA.
  $y$-axis: F1 score.}
    \label{fig:uda_n}
\end{figure}

\newparagraph{Sentence-Level Enrichment.}
From \Cref{tb:combined_result}, we observe that UDA method further enhances the performance of media bias detection, especially in lexical bias detection.
To further explore how UDA functions, we conduct experiments in several steps with the increase in the percentage of unlabeled sentences feeding to the model, and the results of each task are shown in \Cref{fig:uda_n}.

As the scale of training sentences increases, the performance of Coarse-Grained Combined grows continuously with small fluctuations. While the situations are quite different in Fine-Grained classification tasks.
The F1 score of the Fine-Grained lexical bias task first rapidly drops then increases slowly, and maintains an upward trend.
While there is a lot of fluctuation in the curve of the Fine-Grained informational bias task.
The results indicate that enrichment in training corpus has a significant effect on lexical bias detection however the improvements in informational bias are unstable, which is also aligned with their natural features.

When go through the details of the predictions in the model, we observe the UDA's improvements in overfitting. 
It is more sensitive about the biased sentence and can better distinguish the trigger words.
For example, most of the sentences with \textit{Opinion} have the token ''believe'', while the word can also be used for other purposes. UDA can avoid labeling all the sentences with ''believe'' as biased sentences.
Our experiments also verify that large-scale of unlabeled data can be used to enlarge the training set of some practical tasks with small-scale datasets and get improvements in the performance. Also, semi-supervised and even unsupervised learning has the potential to be further explored in the future.

\subsection{Results of RACE}

\label{sec:irmethods}

\newparagraph{Information Retrieval}
Our RACE results are also shown in \Cref{tb:combined_result}.
From our ground truth model BERT+Context, adding external context sentences improves the results, which is aligned with the result of previous work \cite{van2020context}.
Some information retrieval methods are conducted to select valuable information from external knowledge source.
We can observe that the gap between traditional and neural network methods is not as large as other tasks. But the traditional method BM25 is weak in informational bias detection, deep learning models can make up and have more potential in this field.
SimCSE and DPR have similar performance in informational bias performance. While the SimCSE model outperforms other models in lexical bias detection, because we fine-tune the SimCSE model based on a large pre-training language model, which already contains a lot of semantic information.

\newparagraph{Combining Titles}
When the titles are also added as the context of the sentence, from \Cref{tb:combined_result}, we notice that DPR+title gets best performance in general. Compared with the model only consider the single sentence itself, the combination of titles helps to retrieval more precise and valuable information.
While the way to feeding titles is also important, concatenating the title and the sentence simply even sometimes hurts the performances.
DPR model is more flexible, which can encode the titles and sentences separately, takes full advantage of the titles and further improves the results.

We also notice the shortage of our model in considering time series. For example, there is a case that the target sentence is, 

\textit{``If the authorities still further restrict demonstrations and gatherings, they may have exceeded public health safety considerations."} 

We find the context sentences through model DPR+title, for example,

\textit{``Egypt opened the Rafah port connecting the Gaza Strip on the same day, allowing hundreds of Palestinians stranded in Egypt due to the epidemic to return to Palestine"}, 

\textit{``From May 16th to 19th, the ``stay-at-home order" will continue to be implemented nationwide"}, 

\textit{``The spread of the epidemic in India is accelerating"}.

These sentences are similar to the target sentence in semantics, and can also provide information on epidemic situations of other countries as references. However, they will also bring confusion if the description is not for the same period.
The original design of DPR is for the QA system, most of the queries and answers are about general knowledge which seldom changes as time goes on.   
While one of the significant differences between our task and the QA task is that reports in journalism are time-sensitive, which indicates us the future direction of improvement in the models.

\subsection{Case Study}
% We observe that the data augmentation methods work better than informational retrieval methods.
% When analyzing the cases, we have several findings. 

% One of the significant observations is that UDA performs better in fine-grained classification tasks.

The adding of retrieved information provides more knowledge and improves the performance is easy to understand. In this part, to provide more insights into what can be learned by UDA, we conduct a case study on fine-grained lexical bias detection to explore the difference between basic BERT and UDA when detecting the bias after a larger scale of training data is supplied. 

Because there is attention mechanism in both methods, we use the visualization tool BertViz \cite{vig2019multiscale} to show the attention relationship in each layer and head among the tokens.
When we highlight the words according to the attention matrix, we can find that the basic BERT can also learn the pattern related to media bias in general, only minor difference appears in UDA, which improves the final performances.

\begin{figure}[!t]
    \centering 
    \subfigure[BERT]{  
    \begin{minipage}{0.22\textwidth}
        \centering    
        \includegraphics[scale=0.62]{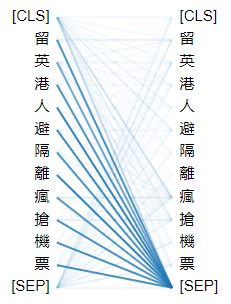}
    \end{minipage}}
    \subfigure[UDA]{
    \begin{minipage}{0.22\textwidth}
        \centering    
        \includegraphics[scale=0.62]{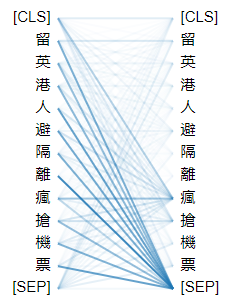}
    \end{minipage}}
    \caption{The attention relationship presentation of model BERT and UDA. The darker color stands for larger value in attention parameters.}  
    \label{fig:casestudy}
\end{figure}

\Cref{fig:casestudy} shows the attention mechanism of the sentence "Hong Kong people who are in the UK are crazy for the air tickets to avoid quarantine". The ground truth shows that this sentence has lexical bias \textit{Exaggeration}.
Both of the methods correctly classify the bias at coarse-grained level, while in fine-grained, basic BERT wrongly classifies it as \textit{Opinion}.
From \Cref{fig:casestudy}, we obverse that the allocation of attention in basic BERT is more average than that in UDA.
As we know, in our dataset, informational bias is longer than lexical bias, and \textit{Opinion} is usually longer than other bias in lexical bias.
Due to the divergence in length of different bias types, it is easier for the model to give the result according to the length of the highlighted part, that is why BERT labels it as \textit{Opinion}.
We notice that in the result of model UDA, the token ''crazy'', which is the trigger of bias \textit{Exaggeration} in human judgment, got much more attention than other tokens.
So, when the model cannot exactly find the important tokens may result in an incorrect result in fine-grained level, and UDA, which enrich data-augmented context and make up the gap.

\section{Related Work}

\subsection{Media Bias Detection}

Our work is in line with prior research in media bias.
Its automatic detection has been examined through substantial efforts with NLP methods.
Most previous studies concern what linguistic features, such as tokens and phrases, will likely be used in news expressing dissenting political opinions \cite{yano2010shedding,baumer2015testing}.
Later work \cite{fan2019plain,van2020context,chen2020detecting} points out the limitation to focus only on lexical bias and calls for attention over information bias.
They integrate some prior efforts in news framing \cite{card2015media,baumer2015testing}, aiming to learn news reporting perspectives, e.g., economy and society as prototype of informational bias, and propose that two types of bias should be considered together.
However, the existing benchmark datasets are mostly labeled in a coarse granularity \cite{lim2020annotating} or through political spectrum \cite{fan2019plain} according to the western ideology, which is not applicable to Chinese media bias detection.
We step further to design fine-grained labels and annotate them in both token and sentence level, and be the first to systematically analysis the media bias from the Chinese perspective.
Also, limited attention has been paid to media bias in non-political news, whereas our paper about data in health field attempts to mitigate the gap.

\subsection{Context Enrichment}
Previous work points out that insufficient data in media bias detection tasks and more context is needed in detecting information bias \cite{van2020context}, while we systematically propose two main directions that enrich the context to help the bias detection tasks. 

\newparagraph{Data-Augmented Context Enrichment}
Collecting a large and well-labeled dataset is expensive, so data augmentation approaches are widely applied in training deep learning models \cite{chen2023empirical}.
Preliminary work focuses on supervised learning tasks, most of which augmentation is at token-level. For example, synonym replacement \cite{kolomiyets2011model}, random deletion \cite{iyyer2015deep}. 
Our work follows Easy Data Augmentation \cite{wei2019eda}, which consists of four basic data augmentation methods, and analyzes the affection of each method.
Later work attempts to augment data at the sentence level directly, for example, Yu et al \cite{yu2018qanet} proposes to enrich training data by paraphrasing. 
As the techniques push on, augmentation data can be used for semi-supervised learning for unlabeled data through consistency training \cite{xie2020unsupervised}. 
We are the first to apply this concept in the media bias detection task.

\newparagraph{Retrieval-Augmented Context Enrichment}
Information Retrieval is widely used as the retriever module in question-answering and fact-checking tasks, which need the support of large knowledge sources \cite{guo2022survey, zhu2021retrieving}.
Traditional methods, such as TF-IDF \cite{kratzwald2018adaptive} and BM25 \cite{yang2019end,wang2019multi}, use sparse representations. Along with the success of deep learning, deep retriever models with dense representations have been developed. 
SimCSE \cite{gao2021simcse} is a simple contrastive learning framework, which makes the distance of the sentences that have similar semantics closer through embedding training. DPR \cite{karpukhin2020dense} trains different encoders to encode the question and document and then measures them by similarity functions.
Based on the frameworks, we propose methods to retrieve more valuable information as context sentences of the target sentences, which largely improve the performance of media bias detection tasks.

\section{Conclusion}

Based on the media bias detection task, we refine the previous label structure and design a new annotation system from the Chinese perspective. We also contribute the first annotated Chinese media bias dataset in the public health field and its benchmark based on a multi-label setting.
In a further step, we explore how context enrichment methods can be better applied to the practice problem with small-scale datasets.
Through substantial experiments on it, we point out these crucial findings: 
(1) Training data scale may largely limit the model's capability to learn media bias, and contextual enrichment in both data augmentation and information retrieval is potentially helpful.
(2) Retrieval-augmented methods work better than data-augmented methods, especially when more context information, such as the titles' feature, is added. However, there are also some details worth further exploring in future work.

%% The Appendices part is started with the command \appendix;
%% appendix sections are then done as normal sections
%% \appendix

%% \section{}
%% \label{}
\appendix
\setcounter{table}{0}
\renewcommand{\thetable}{\thesection.\arabic{table}}

\setcounter{figure}{0}    
\renewcommand{\thefigure}{\thesection.\arabic{figure}}

\section{Original Text of Examples}

% Due to the length limitation of our paper, we show original Chinese text of \Cref{fig:intro} and \Cref{tb:example} in \Cref{tb:appendix_intro} and \Cref{tb:appendix_example}.

% \input{tables/appendix_intro.tex}
% \input{tables/appendix_example.tex}

\begin{table*}[htbp]
\centering
\small
\begin{tabular}{p{0.85\textwidth}}
\toprule
\textit{\underline{Article 1}}
“Banquet for tens of thousands” coursed the trouble? Wuhan community $[$explodes with$]$ collective fever. …
\\
\begin{CJK*}{UTF8}{bsmi}
萬家宴惹禍？武漢社區$[$爆$]$集體發熱。……
\end{CJK*}\\
$[$\underline{Experts} pointed out that the Banquet has exceeded the 2-week incubation period so far, “Those who currently have fever symptoms may be in contact with coronary pneumonia patients at a later stage”.$]$ …
\\
\begin{CJK*}{UTF8}{bsmi}
$[$\underline{專家}表示，萬家宴於1月18日舉行，至今已經超過2周潛伏期，「目前有發熱症狀的，有可能是後期接觸了新冠肺炎的病人」。$]$ ……
\end{CJK*}
\\
$[$\underline{Secretary Gong of the Party Branch of Kanghe Neighborhood Committee in Baibuting} pointed out that the main purpose for labeling the “hot building” is to remind residents to avoid going out as much as possible.$]$ …
\\
\begin{CJK*}{UTF8}{bsmi}
$[$\underline{百步亭社區康和居委會黨支部龔書記}指出，發熱門棟主要是為提醒居民，盡可能避免出門。$]$ ……
\end{CJK*}
\\\midrule
\textit{\underline{Article 2}}
“Banquet for tens of thousands” was still held before lunar New Year, Wuhan Baibuting was $[$seriously infected and despairingly crying$]$. …
\\
\begin{CJK*}{UTF8}{bsmi}
年前仍擺萬家宴，武漢百步亭$[$感染嚴重哭喊絕望$]$。……
\end{CJK*}\\
$[$A \underline{netizen} posted on Sina Weibo late at night on the 9th, saying that he was writing this in “desperation” and that Baibuting is currently “in an uncontrolled situation”. Because of the banquet, many people were infected with pneumonia.$]$ …
\\
\begin{CJK*}{UTF8}{bsmi}
$[$\underline{網友}9日深夜在新浪微博發文，指自己是在「絕望中寫這些」，目前百步亭處於「無人管的境地」。 因為舉辦「萬家宴」，導致很多人感染了武漢肺炎。$]$……
\end{CJK*}
\\
This post was widely forwarded more than 100,000 times on Sina Weibo, $[$but \underline{Wuhan and even Chinese officials} have yet to respond to the allegations in the post.$]$ …
\\
\begin{CJK*}{UTF8}{bsmi}
這篇貼文發出後，在新浪微博被廣為轉發超過10萬次，$[$但\underline{武漢乃至於中國官方}，目前尚未對貼文中指控的內容作出回應。$]$ ……
\end{CJK*}
\\
\bottomrule
\end{tabular}
\caption{The original text of examples in \Cref{fig:intro}.}
\label{tb:appendix_intro}
\end{table*}

\begin{table*}[htbp]
\centering
\small
\begin{tabular}{p{0.97\textwidth}}
\toprule
\textbf{Writing Bias Annotation Examples} \\\midrule
(a) ... They are $[$crazy for$]_\textbf{Exaggeration}$ air tickets, flying around the earth, hoping to "return to $[$the safest$]_\textbf{Exaggeration}$ place" ... \\
\begin{CJK*}{UTF8}{bsmi}
……他們$[$瘋搶機票$]$, 繞著地球飛, 盼「回到$[$最安全$]$的地方」……
\end{CJK*}\\\midrule

(b) ... some people even continue to go out to have dinner with friends. The ``$[$Buddhism$]_\textbf{Stereotype/Labeling}$'' is surprising ...\\
\begin{CJK*}{UTF8}{bsmi}
……甚至還有人繼續在外和朋友聚餐等，$[$佛系$]$得讓人意外。
\end{CJK*}\\\midrule

(c) ... $[$some experts doubt the effectiveness of mass quarantine, pointing out that ``the virus will actually spread''$]_\textbf{Ventriloquism}$\\
\begin{CJK*}{UTF8}{bsmi}
……$[$有專家懷疑大規模隔離的成效，指「病毒實際上總會散播開去」$]$
\end{CJK*}\\\midrule

(d) ... and the situation $[$is bound to continue to deteriorate$]_\textbf{Speculative Content / Opinion}$ ...\\
\begin{CJK*}{UTF8}{bsmi}
……情況$[$想必會持續惡化$]$……
\end{CJK*}\\

\bottomrule
\toprule
\textbf{Content Bias Annotation Examples}
\\\midrule
\textbf{\underline{Annotation Set 1}}\\
\textit{\underline{Article 1}}
... The Chinese song made many $[$\underline{foreign netizens} sarcastically say in a ``high-level black'' way that this ``North Korean boy'' is too cute.$]$…\\
\begin{CJK*}{UTF8}{bsmi}
中國歌曲讓不少[外國網友以「高級黑」方式諷刺說，這「北韓男孩」太可愛了。
\end{CJK*}\\
$[$According to \underline{``The Paper''}, the song has a detailed and in-depth description of the mobile cabin hospital, which helps patients fight the virus and relieve tension. $]_\textbf{Framing / Undue Weight (Positive)}$ ...\\
\begin{CJK*}{UTF8}{bsmi}
據《澎湃新聞網》聲稱，該首歌曲對於方艙醫院有細緻、深入的具體刻畫，有助於患者對抗病毒、舒緩緊張
\end{CJK*}
\\
\specialrule{0.00001em}{1pt}{1pt}
\textit{\underline{Article 2}}
... $[$\underline{Some netizens} denounced the creator for ``entertaining disasters''$]$. $[$\underline{The Beijing News} online commentary even denounced this as ``forcibly extolling disasters''. $]_\textbf{Framing / Undue Weight (Negative)}$  ...\\
\begin{CJK*}{UTF8}{bsmi}
有網民直斥創作人「把災難娛樂化」]，[《新京報》網上評論更直斥這是「強行歌頌災難」。
\end{CJK*}\\
$[$\underline{Jiang Junrong} said ``A lot of places are `cleared' now. Don't we need to show a little bit of optimism? Don't we need to show a better mentality?'' $]_\textbf{Omission of Article 1}$ ...\\
\begin{CJK*}{UTF8}{bsmi}
蔣軍榮接受澎湃新聞訪問時表示，「現在好多地方都『清零』了。我們不需要拿出一點樂觀主義精神嗎？ 不需要拿出更好的心態來嗎？」 
\end{CJK*}
\\
\specialrule{0.00001em}{1pt}{1pt}
\textit{\underline{Article 3}}
... $[$\underline{Some netizens} even described the video as ``seemingly back in the 1980s and 1990s''.$]$ ...\\
\begin{CJK*}{UTF8}{bsmi}
有網民甚至形容視頻“好像回到了八九十年代”。
\end{CJK*}\\
$[$Composer \underline{Jiang Junrong} said in response that ``They don't know my state of life, and I don't make any calculations''.$]_\textbf{Omission of Article 1}$\\
\begin{CJK*}{UTF8}{bsmi}
作曲家蔣軍榮回應時說：“他們不知道我的生命狀態，我也不會計較。” 
\end{CJK*}\\
$[$\underline{The Beijing News} commented on this, saying that turning it into the material of cheerful children's songs not only dilutes the disaster background, but also eliminates the seriousness of the fight against the epidemic. $]_\textbf{Framing / Undue Weight (Negative)}$ \\
\begin{CJK*}{UTF8}{bsmi}
《新京報》對此發表評論文章認為，中國抗擊冠病疫情是一件嚴肅而沉重的事情，將方艙醫院化為歡快兒歌的素材，不僅淡化了抗疫的災難底色，也消解了抗疫的嚴肅性。
\end{CJK*}
\\\midrule
\textbf{\underline{Annotation Set 2}}\\
\textit{\underline{Article 1}}
... ``Shincheonji'' chairman will hold a press conference to publicly respond to the epidemic for the first time.\\
\begin{CJK*}{UTF8}{bsmi}
韓國「新天地」會長將開記者會，首次公開回應疫情。
\end{CJK*}
\\
\textit{\underline{Article 2}}
South Korean leader of ``Shincheonji'' kneels to apologize and thank the government for fighting the epidemic.\\
\begin{CJK*}{UTF8}{bsmi}
南韓新天地教主下跪道歉，又感謝政府抗疫。
\end{CJK*}
\\
\textit{\underline{Article 3}}
Weird! The leader of ``Shincheonji'' kneels and $[$wears Park Geun-Hye's watch$]_\textbf{Sensationalism Content}$.\\
\begin{CJK*}{UTF8}{bsmi}
詭！新天地教主下跪，竟戴著朴槿惠手錶。
\end{CJK*}
\\
\bottomrule
\end{tabular}
\caption{The original text of examples in \Cref{tb:appendix_intro}. }
\label{tb:appendix_example}
\end{table*}
Due to the length limitation of our paper, we show original Chinese text of \Cref{fig:intro} in \Cref{tb:appendix_intro}, and \cref{tb:example} in \Cref{tb:appendix_example}.

% % \input{tables/appendix_example.tex}

% \begin{figure*}[htbp]
%     \centering
%     \includegraphics[width=0.85\textwidth]{pictures/tool_content.png}
%     \caption{The interface of content bias annotation tool, which has the same functions with writing bias annotation tool. Several questions are set to assist the determination of annotators. }
%     % \vspace{-.6cm}
%     \label{fig:tool_content}
% \end{figure*}

\section{Annotation Interface}

\begin{figure*}[htbp]
    \centering
    \includegraphics[width=0.85\textwidth]{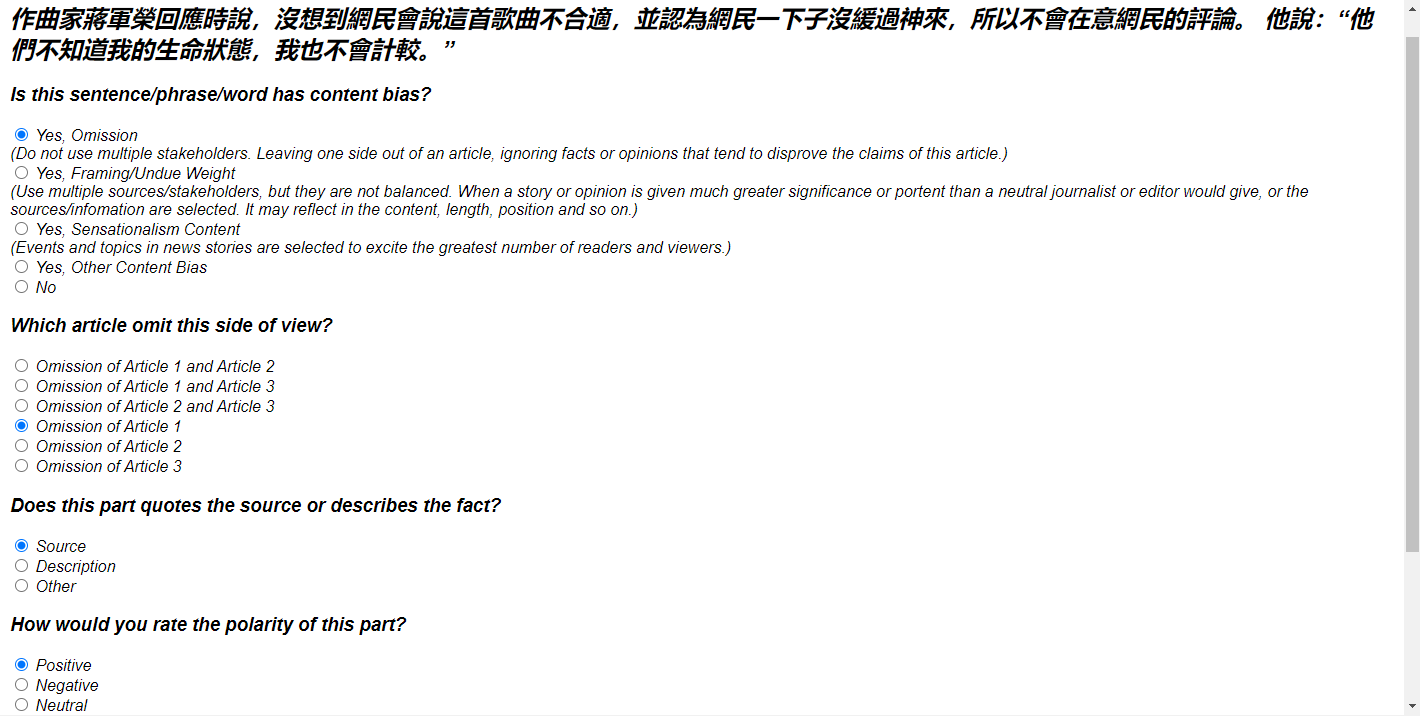}
    \caption{The interface of content bias annotation tool, which has the same functions with writing bias annotation tool. Several questions are set to assist the determination of annotators. }
    % \vspace{-.6cm}
    \label{fig:tool_content}
\end{figure*}

% In our dataset, the annotation works in writing bias (\Cref{fig:tool_writing}) and content bias (\Cref{fig:tool_content}) are conducted separately, while the basic functions of the tools are the same.
Interface of content bias annotation tool is shown in \Cref{fig:tool_content}.

% \begin{figure*}[htbp]
%     \centering
%     \includegraphics[width=0.85\textwidth]{pictures/tool_writing.png}
%     \caption{The interface of writing bias annotation tool. Except for the sentence-level annotation, the annotators can also annotate in spans. }
%     % \vspace{-.6cm}
%     \label{fig:tool_writing}
% \end{figure*}

% \clearpage

\section{GPT Prompts}
\label{app:prompts}

We show the prompt of Fine-Grained Combined bias detection task as the example:\\

% \shi{make the font of the prompt a smaller size}
\scriptsize\textit{
Given the fine-grained class definitions of media bias as following (Exaggeration, Stereotype, Ventriloquism, Opinion, Omission, Framing, Sensationalist):\\
Exaggeration(It is defined as using eye-catching wordings with strong emotional implications or describing things in a way that is more than it really is, to influence an audience.),\\
Stereotype(Sometimes a media source may favor or attack a particular group, or a person, and describe them with wordings reflecting their stereotype to them. We refer to this kind of bias as Stereotype, or we can call it labeling. It can also describe the events or people using words that may contain judgments.), \\
Ventriloquism(We annotate text spans as Ventriloquism when experts or witnesses are quoted in a way that intentionally voices the authors own opinion),\\
Opinion(This sub-type covers cases like stories that focus not on what has occurred, but primarily on what might occur, or it is just the opinion of the author. We can also name it Speculative Content.),\\
Omission(We use Omission to describe situations where some reports ignore information that tends to disprove their claims: if an article does not contain the content about stakeholders that appeared in other articles, then it has the informational bias of Omission.),\\
Framing(We say an article has the informational bias of Framing when multiple stakeholders are contained but not balanced or undue weight. Such imbalances can be raised from the following aspects: different positions, different lengths of the description, and different sentiment polarities of selected content.), \\
Sensationalist(This informational bias sub-type refers to the event that is not closely relevant to the theme, but selected to excite the greatest number of readers.)\\
 -----------------------------\\
Given the sentence: SENTENCE\\
Please answer the informational bias situation of this sentence.\\
Format: $<$Exaggeration (If Yes: 1, else: 0)$>$,$<$Stereotype (If Yes: 1, else: 0)$>$,$<$Ventriloquism (If Yes: 1, else: 0)$>$,$<$Opinion (If Yes: 1, else: 0)$>$,$<$Omission (If Yes: 1, else: 0)$>$,$<$Framing (If Yes: 1, else: 0)$>$,$<$Sensationalist (If Yes: 1, else: 0)$>$\\
Please only answer the number.
}

%% If you have bibdatabase file and want bibtex to generate the
%% bibitems, please use
%%
% \bibliographystyle{} 
\bibliographystyle{elsarticle-num} 
\bibliography{custom}

%% else use the following coding to input the bibitems directly in the
%% TeX file.

% \begin{thebibliography}{00}

% %% \bibitem{label}
% %% Text of bibliographic item

% \bibitem{}

% \end{thebibliography}

% \printbibliography

\end{document}